# CCD photometry and visual observations of V1663 Aquilae (Nova Aquilae 2005)

**David Boyd & Gary Poyner**

We present CCD V and I band photometry and visual observations of V1663 Aquilae (Nova Aquilae 2005). This was a classical fast nova, probably of type Ba, observed for the first time on 2005 June 9. Maximum light occurred at HJD= 2453531.2 ± 0.2, when the apparent V magnitude was 10.7 ± 0.1 and the V–I colour index 3.1. Decline times were $t_{2,V}$= 17d, $t_{2,I}$= 18d, $t_{3,V}$= 32d and $t_{3,I}$= 35d. We derived a maximum absolute V magnitude of −7.8 ± 0.2, colour excess E(B−V)= 2, extinction $A_V$= 6 magnitudes and distance d= 2.9 ± 0.4 kpc. After maximum the V–I colour index remained between 3.1 and 3.4 for 50 days, then gradually reduced as the nova became bluer with increasing age. We saw no direct evidence of dust emission although this could have contributed to the high extinction in the direction of the nova.

## Introduction

Novae occur in close binary systems in which material is being transferred from a cool secondary star to a more massive white dwarf primary. Hydrogen-rich gas flows from the secondary into the accretion disc round the white dwarf and then down onto its surface. Here it builds up until it reaches a sufficient temperature and density that a runaway thermonuclear reaction starts at the base of the hydrogen layer. The resulting explosion produces a sudden large increase in brightness of the system and blows the surface of the white dwarf away as an expanding shell of hot gas. As this gas slowly cools the light output of the system gradually declines.

Novae come in four varieties: fast novae which rise steeply to a maximum and then decay by around three magnitudes in less than 100 days; slow novae which rise more gradually and decline in more than 100 days; very slow novae, often called symbiotic novae, which remain near maximum light for many years and then decay very slowly; and recurrent novae, which undergo repeated outbursts at time intervals of many years. Fast and slow novae are generally referred to as classical novae.

## Discovery

Nova Aquilae 2005 was first reported by Grzegorz Pojmanski on behalf of the All Sky Automated Survey (ASAS) on 2005 June 10.[1] An image taken on 2005 June 9.240 (HJD=2453530.7445) recorded a new object at RA 19h 05m 12s, Dec +05d 14m 12s with magnitude V=11.046. No object brighter than V=14 was present at this position on an image taken on 2005 June 3.318. The object was confirmed by Arto Oksanen at V=10.85 on 2005 June 10.361 (HJD=2453531.8655).

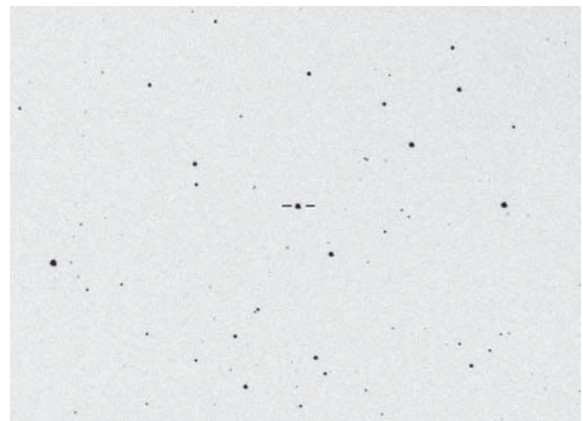

**Figure 1.** Field of V1663 Aquilae (N up, E to left, width 15'). *David Boyd*

Official announcement of the discovery on IAU *Circular* 8544 assigned the designation V1663 Aquilae.

Gary Poyner (GP) made his first visual estimate of v=11.3 on June 10.986. David Boyd (DB) made his first CCD measurements of V=11.22 and I=8.11 on June 13.943. Both observers were using 0.35m SCTs. DB used an SXV-H9 CCD camera with V and I-band filters. Exposures were increased from 8sec near maximum to 60sec as the nova faded. The V band image in Figure 1 taken on June 17.976 shows the field of the nova. Astrometry using *Astrometrica*[2] and the USNO *CCD Astrograph Catalog* release 2 (UCAC2)[3] gave the position RA 19h 05m 12.50±0.09s, Dec 05d 14m 11.45±0.13s (2000). Galactic coordinates are l= 39°.2, b= −0°.7.

## Establishing comparison stars

Before the brightness of the nova could be measured, comparison stars had to be established in the field of the nova. A chart was published by Reinder Bouma on 2005 June 11 based on ASAS V band measurements which gave comparison magnitudes accurate to 0.1 magnitude suitable for making visual estimates.[4]

However, for CCD photometry, V and I band comparison magnitudes were needed to higher precision. This was achieved by DB using 5 nearby *Hipparcos* stars (93274, 93606, 93797, 93856, 94139), for which Vj and (V–I)c magnitudes were available, as secondary standards to determine V and I magnitudes of 5 of the comparison stars on the Bouma chart (110, 116, 123, 124 and 131). On a good quality night, each of the *Hipparcos* stars was measured 10 times in V and I and the means of these measurements were used to define local V and I zero points. Experience has shown that by using an ensemble of 5 or 6 close *Hipparcos* stars, local zero points can be determined to better than 0.008 magnitude in V and 0.03 magnitude in I. Using these local zero points, and applying transformations to the standard Johnson–Cousins photometric system, we derived V and I magnitudes of the 5 comparison stars. The absolute accuracy of this calibration is estimated to be better than 0.015 magnitude in V and 0.05 magnitude in I. These 5 comparison stars were then used as an ensemble to determine V and I magnitudes of the nova throughout the outburst.

Each magnitude measurement of the nova was a mean from at least 5 images in each passband, and all measurements were transformed to the standard Johnson–Cousins system. As an indication of consistency over the 6-month period of these observations, the spread of the magnitude of each comparison star within the ensemble was, on average, 0.005 in V and 0.008 in I. After 50 days, the brightest comparison star was dropped from the ensemble when the exposures were increased to compensate for the reduction in brightness of the nova.

## Observations

Figure 2 shows the visual and CCD magnitude measurements and Figure 3 the V–I colour index. Measurement errors are plotted but in most cases are smaller than the drawn symbols. The visual estimates closely followed the CCD V measurements but were, on average, 0.2 to 0.3 magnitude fainter. After 80 days the nova was too faint for reliable visual estimates. After 180 days the nova was inaccessible to the CCD because of local obstructions.

## Analysis

### Decline parameters

Decline rates of novae are characterised by the light curve decline parameters $t_2$ and $t_3$ which are the times taken to drop by 2 and 3 magnitudes below maximum light.

Considering the distribution of observations around the time of maximum light, we conclude that maximum light probably occurred at HJD= 2453531.2 ± 0.2, at apparent V and I magnitudes $m_V$(max)= 10.7 ± 0.1 and $m_I$(max)= 7.6 ± 0.1 (assuming V–I= 3.1 at maximum).

We fitted 6th order polynomials to the CCD V and I observations to achieve a close fit to the data in the important early part of the light curves. There were sufficient data points to well-constrain these high order fits. Lower order polynomials did not fit the early part of the light curves well. Using these fitted polynomials, we determined the decline parameters to be:

$$t_{2,V}= 17d \quad t_{2,I}= 18d$$
$$t_{3,V}= 32d \quad t_{3,I}= 35d$$

Taking into account the difficulty in determining the time and magnitude of maximum, the uncertainty in these values is ± 2d. The same analysis of the visual observations gave similar decline times as were obtained for the CCD V data. The corresponding decline rate $v_{d2}$ is 0.11±0.01 magnitude/day. This makes V1663 Aquilae a classical fast nova with a decline rate typical for a nova in our galaxy.

Novae light curves have been classified by Duerbeck[5] according to their shape and decline times. By his criteria, given the shape of the light curve and the observed value of $t_3$, V1663 Aquilae is most probably a type Ba nova. Duerbeck's description of type Ba is 'decline with standstills or other minor irregular fluctuations during decline'. Spectroscopic observations would be needed to confirm this assignment.

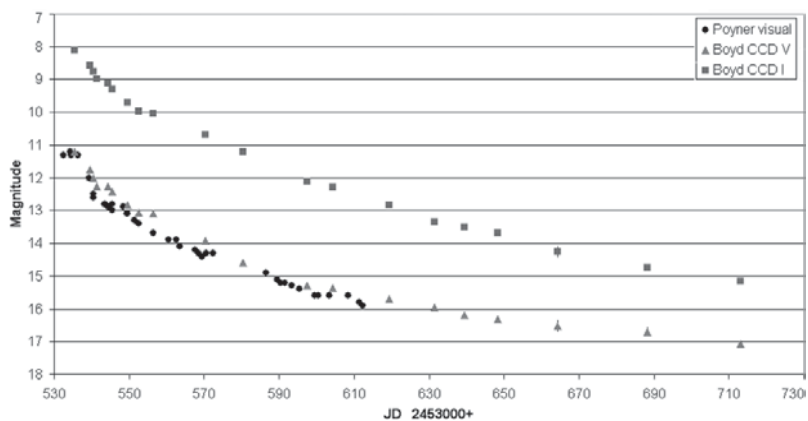

**Figure 2.** Visual, CCD V and CCD I light curves of V1663 Aquilae.

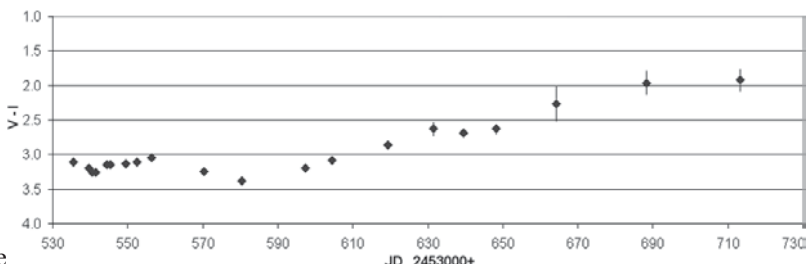

**Figure 3.** V–I colour index of V1663 Aquilae. *David Boyd*

Table 1. Maximum absolute V magnitude of V1663 Aquilae

| Method | $M_V(max)$ |
|---|---|
| Linear model for galactic novae[6] | −7.7±0.6 |
| S-shaped model for extragalactic novae[7] | −8.2±0.6 |
| Linear model for novae of types B, C, or D[8] | −7.4±0.6 |
| Assuming constant $M_V(15)$[8] | −7.9±0.5 |

*Maximum absolute magnitude*

The relationship of maximum magnitude to rate of decline has been extensively researched in order to provide an additional galactic distance estimation tool. This has been calibrated using novae in our own galaxy, in M31 and in the Magellanic Clouds. Various formulae for this relationship have been given by, among others, Cohen,[6] Della Valle & Livio[7] and Downes & Duerbeck.[8] Taking our measured value of $t_{2,V}$, we used these formulae to derive values for the maximum absolute V magnitude of the nova, $M_V(max)$. Buscombe & de Vaucouleurs[9] found that the absolute V magnitudes of novae 15 days after maximum light, $M_V(15)$, were approximately constant. Subsequent authors have confirmed this relationship but have published different constant values. Downes & Duerbeck,[8] using a sample of 28 objects, gave $M_V(15) = -6.05 \pm 0.44$. We measured the apparent V magnitude of V1663 Aql after 15 days to be $m_V(15) = 12.5 \pm 0.1$, and knowing $m_V(max)$ could calculate $M_V(max)$.

The values of $M_V(max)$ produced by these four different methods are listed in Table 1. The errors given include the errors associated with the formulae plus the uncertainties in our value of $t_{2,V}$ and in magnitude determinations. The mean value for $M_V(max)$, the maximum absolute V magnitude of V1663 Aquilae, is −7.8 with a standard error of 0.2.

*Colour excess, extinction and distance*

When first measured 4 days after the outburst, the V−I colour index was 3.1. It varied between 3.1 and 3.4 for 50 days before reducing gradually to 1.9 after 180 days, following the normal tendency of novae to become bluer as they age.

According to van den Bergh & Younger,[10] the mean intrinsic colour index $(B-V)_0$ of galactic novae at time $t_{2,V}$ is −0.02±0.04. We assume that their intrinsic colour index $(V-I)_0$ at time $t_{2,V}$ is also close to 0.0, and since the apparent (V−I) colour index of V1663 Aquilae at this time is 3.1, we estimate the colour excess E(V−I) of the nova to be 3.1. Savage and Mathis[11] give the interstellar colour excess ratio E(V−I)/E(B−V) as 1.6, which implies a colour excess E(B−V)= 2. The interstellar ratio of total to selective extinction, $R_V$, is conventionally taken as 3.1.[11] Using the relationship $A_V = R_V * E(B-V)$, we estimate $A_V$, the total interstellar extinction in V, to be 6 magnitudes. Using the distance modulus corrected for extinction

$$m_V(max) - M_V(max) = 5 \log d - 5 + A_V$$

we determine d, the distance to the nova, to be 2.9 ± 0.4 kpc.

We consulted measurements of interstellar extinction published by Neckel & Klare[12] to investigate expected absorption at the galactic coordinates of the nova (l= 39°.2, b= −0°.7). While they present no data in exactly this direction, the nearest region they analyse has extinction of ~3 magnitudes within 1kpc but no measurement of extinction beyond that distance. This is consistent with our measurements of extinction and distance. It is possible that additional extinction could be caused by the emission of dust following the nova outburst, as was postulated in V475 Scuti.[13] In that case the emission of dust was associated with a rapid drop of 4 magnitudes in V and a simultaneous increase in the V−I colour index from 1.5 to 3, both of which took place around 60 days after maximum. No such behaviour was observed in V1663 Aquilae. There was a long steady decline in visual brightness and the V−I colour index remained between 3.1 and 3.4 for 50 days before decreasing slowly. We conclude that we did not see a similar indication of dust emission in V1663 Aquilae. In summary, therefore, it is likely that most of the reddening we observed was due to interstellar extinction, although dust emitted by the nova may have contributed.

## Summary

V1663 Aquilae (Nova Aquilae 2005), observed for the first time on 2005 June 9, was a classical fast nova, probably of type Ba. Its subsequent decline was observed for 80 days visually and for 180 days photometrically in V and I using a CCD. We estimate maximum light occurred at HJD= 2453531.2 ± 0.2 when the apparent V magnitude was 10.7±0.1 and the V−I colour index 3.1. Decline times were $t_{2,V}$= 17d, $t_{2,I}$= 18d, $t_{3,V}$= 32d and $t_{3,I}$= 35d. We derived a maximum absolute V magnitude of −7.8±0.2, colour excess E(B−V)= 2, extinction $A_V$= 6 magnitudes and distance d= 2.9±0.4 kpc. After maximum the V−I colour index remained between 3.1 and 3.4 for 50 days and then gradually reduced as the nova became bluer. We saw no direct evidence of dust emission although this could have contributed to the high absorption in the direction of the nova.

## Acknowledgments

The authors thank the referees, Drs Chris Lloyd, Darren Baskill and Jeremy Shears for their constructive comments which have helped us to improve the paper.

**Addresses: DB:** 5 Silver Lane, West Challow, Wantage, OX12 9TX. [drsboyd@dsl.pipex.com]
**GP:** 67 Ellerton Road, Kingstanding, Birmingham B44 0QE. [garypoyner@blueyonder.co.uk]